\documentclass[a4paper]{jpconf}
\usepackage{bm}
\usepackage{graphicx,subfigure,xcolor}
\usepackage{braket}

\def\xt{x(\tau)}
\def\xtp{x(\tau')}
\def\xAt{x_A(\tau)}

\def\xBt{x_B(\tau)}
\def\xBtp{x_B(\tau')}
\def\w{\omega}
\def\w0{\omega_0}

\def\bk{{\bf k}}

\def\br{{\bf r}}
\def\bR{{\bf R}}

\def\vac{\mid 0\rangle}
\def\vacd{\langle 0\mid}

\def\bmu{{\boldsymbol \mu}}

\begin{document}

\title{Van der Waals and resonance interactions between accelerated atoms in vacuum and the Unruh effect}

\author{M Lattuca$^1$, J Marino$^2$, A Noto$^1$, R Passante$^{1,3}$, L Rizzuto$^{1,3}$,\\
S Spagnolo$^1$, W Zhou$^{1,4}$}

\address{$^1$ Dipartimento di Fisica e Chimica, Universit\`{a} degli Studi di Palermo, Via Archirafi 36, 90123 Palermo, Italy\\
$^2$ Institut fur Theoretische Physik, Universit\"{aÌ}t zu Koln, Zulpicher Stra{\ss}e 77, 50937 Cologne, Germany\\
$^3$ INFN, Laboratori Nazionali del Sud, 95123 Catania, Italy\\
$^4$ Center for Nonlinear Science and Department of Physics, Ningbo University, Ningbo, Zhejiang 315211, China}
\ead{roberto.passante@unipa.it}

\begin{abstract}
We discuss different physical effects related to the uniform acceleration of atoms in vacuum, in the framework of quantum electrodynamics. We first investigate the van der Waals/Casimir-Polder dispersion and resonance interactions between two uniformly accelerated atoms in vacuum.
We show that the atomic acceleration significantly affects the van der Waals force, yielding a different scaling of the interaction with the interatomic distance and an explicit time dependence of the interaction energy. We argue how these results could allow for an indirect detection of the Unruh effect through dispersion interactions between atoms.
We then consider the resonance interaction between two accelerated atoms, prepared in a correlated Bell-type state, and interacting with the electromagnetic field in the vacuum state, separating vacuum fluctuations and radiation reaction contributions, both in the free-space and in the presence of a perfectly reflecting plate.
We show that nonthermal effects of acceleration manifest in the resonance interaction, yielding a change of the distance dependence of the resonance interaction energy. This suggests that the equivalence between temperature and acceleration does not apply to all radiative properties of accelerated atoms. To further explore this aspect, we evaluate the resonance interaction between two atoms in non inertial motion in the coaccelerated (Rindler) frame and show that in this case the assumption of an Unruh temperature for the field is not required for a complete equivalence of locally inertial and coaccelerated points of views.
\end{abstract}

\section{Introduction}
A remarkable consequence of quantum field theory in accelerated frames is that the concept of vacuum is observer-dependent \cite{Fulling73,Davies75, Unruh76}. When viewed from a uniformly accelerated observer, the vacuum of a scalar field appears as a {\em thermal} bath of quanta at a temperature proportional to its acceleration, $T_U=\hbar a/(2\pi ck_B)$, $a$ being the proper acceleration and $k_B$ the Boltzmann constant.
This phenomenon, which is the analogous in a flat space-time of the Hawking radiation from a black hole, is known as Davies-Fulling-Unruh effect, and its essence is that the particle content of a quantum field is not invariant for all coordinate transformations,  but it depends on the reference frame \cite{Unruh76,CHM08}.
This conceptually subtle result, on the border between quantum field theory and general relativity,  has been object of intense investigations in the last decades, and some controversial aspects concerning its physical meaning are still debated \cite{FMNBK99,FC06,BS13}. Several experimental schemes to detect the Unruh effect have been proposed in the literature \cite{CHM08,Yablonovitch89,SSH06,CLMV17,RCPR08}, but a direct verification of this effect is still a challenge for experimentalists, due to the very high accelerations necessary to reveal a Unruh temperature of even a few Kelvin.
For example, an acceleration of about $9.8$ m/s$^2$, the gravitational acceleration experienced by a detector near the Earth's surface, gives a Unruh temperature of the order of $10^{-20}$ K, which is very difficult to detect.

Even if the absence of any direct experimental confirmation of the Unruh effect has led some physicists to question the very existence of this effect, it has been pointed out that the Unruh effect is a consequence of quantum field theory necessary to keep its coherency in accelerated systems \cite{MV02}.
A direct or indirect detection of the effect could thus allow to settle a fundamental question in quantum field theory.

A different although related question recently addressed in the literature, is whether the
effects of a relativistic acceleration are strictly equivalent to a thermal field and what is the physical meaning of this equivalence \cite{BS13, MV02}. It has been recently discussed, for example, that nonthermal features associated with a uniform acceleration appear in the radiative properties of single accelerated atoms \cite{Passante98,RS09}, as well as in the van der Waals/Casimir-Polder interaction between atoms in non inertial motion \cite{MNP14}, or in the entanglement generation between two accelerating atoms \cite{HY15,BF04}.

Motivated by these intriguing issues, in this paper we investigate and review the effect of a uniformly accelerated motion on the van der Waals and resonance interactions between two atoms in vacuum,
with the aim to explore different physical manifestations of the Unruh effect, as well as the possibility to observe signatures of  such effect on the radiation-mediated interaction between accelerating atoms.

Van der Waals and Casimir-Polder dispersion forces are long-range interactions between atoms or neutral objects due to zero-point fluctuations of the quantum electromagnetic field and exchange of virtual photons between the atoms \cite{PT93,CT98}.
These interactions, of purely quantum origin, have been measured with accuracy \cite{Lamoreaux05,Buhmann12}.
When one of the two atoms is in an excited state, a real resonant photon can be exchanged between the atoms. These interactions are usually a fourth-order effect in the matter-field coupling, and thus they are very tiny. Resonance interactions can occur at second order in the coupling when two identical atoms are prepared in a correlated (symmetric or antisymmetric) state, with one atom excited and the other in the ground state. In this case the excitation is delocalized over the two atoms. Since such interaction is a second-order effect, it is usually much more intense than dispersion interactions and, moreover, can be of very long range, scaling asymptotically as $R^{-1}$ ($R$ being the interatomic distance) \cite{CT98, Salam10}.

In this paper, we explore the effect of atomic acceleration on both van der Waals/Casimir-Polder and resonance interactions between two atoms moving with the same uniform acceleration in the vacuum, in the free space or in the presence of a reflecting plane boundary.
We show that the atomic acceleration significantly affects the van der Waals interaction between the two accelerated atoms, resulting in a different scaling of the interaction with the distance compared to inertial atoms and in an explicit time dependence of the interaction. When sufficiently long times are taken, we show that the change in the interaction energy due to the acceleration may become significant even for reasonable values of the acceleration \cite{NP13}. This suggests that van der Waals interactions could be used as a probe to obtain an indirect signature of the Unruh effect,  without necessity of extremely high accelerations \cite{NP13}.
We then investigate the effect of atomic acceleration on the resonance interaction between two entangled identical atoms accelerating in the vacuum space. By separating vacuum fluctuations and radiation reaction contributions, we find that
the resonance interaction is exclusively due to the radiation reaction contribution, while vacuum fluctuations do not contribute \cite{RLMNSZP16, ZRP17}. Accordingly, Unruh thermal fluctuations, which are basically related to the vacuum field correlations in the locally inertial frame of both atoms, do not affect the resonance interaction between the accelerated atoms. Neverthless, the features of the interaction changes qualitatively due to acceleration, showing the presence of purely nonthermal effects of acceleration in the resonance interaction (i.e. not related to the equivalence between acceleration and temperature).
In order to further address this point, we have investigated the resonance interaction between accelerated atoms also from the point of view of a coaccelerated frame. We show that, although the vacuum state in the comoving (i.e. locally inertial) and coaccelerated frames is different due to the Unruh effect, the resonance interaction between two atoms as measured by observers in these frames is the same, without the assumption of an Unruh temperature for the quantum field in the comoving frame \cite{ZPR16}. This result is at variance with what has been obtained for other radiative effects, such as the spontaneous emission, where vacuum fluctuations are relevant \cite{ZY07}. Thus we conclude that the necessity to introduce an Unruh temperature in the coaccelerated frame depends on the specific physical process considered. Finally, we briefly discuss how the presence of a perfectly reflecting plate affects the resonance interaction between two accelerating atoms, showing the appearance of new features that could be relevant for detecting signatures of an accelerated motion in quantum field theory.

The paper is structured as follows. In Section \ref{sec2} we investigate the van der Waals interaction between two uniformly accelerating atoms in vacuum. In Section \ref{sec3} we discuss the effect of a uniform acceleration on the resonance interaction between two accelerated atoms in vacuum, both in the comoving and in the coaccelerated frame. Finally, in Section \ref{sec4} we briefly discuss the effect of a perfectly reflecting mirror on the resonance interaction between two accelerated atoms. Last Section is devoted to our conclusive comments.

\section{The van der Waals interaction between two uniformly accelerated atoms in vacuum}
\label{sec2}
Let us consider two atoms, $A$ and $B$, uniformly accelerating in the vacuum with the same acceleration. We assume the acceleration  orthogonal to their distance, so that their separation is constant. The atoms interact with the quantum electromagnetic field in the vacuum state. We wish to evaluate the van der Waals/Casimir-Polder interaction energy between the two accelerating atoms. In order to do this, we exploit the heuristic model in \cite{PT93}, according to which the atom-atom Casimir-Polder interaction originates from the (classical) interaction between the instantaneous dipole moments induced and correlated by the spatially correlated zero-point fluctuations of the quantum electromagnetic field, and extend this model to accelerated atoms. For two ground-state atoms at rest separated by a distance $\bR=\br_A-\br_B$, the Casimir-Polder interaction is thus obtained as
\begin{eqnarray}
\Delta E_{CP}=\sum_{{\bk} j}\alpha_A(k)\alpha_B(k) \vacd E_{\ell}({\bk},j;\br_A)E_{m}({\bk},j;\br_B)\vac(V_{\ell m}(\bk,{\bf R}))
\label{eq:1}
\end{eqnarray}
where $\alpha(k)$ is the atomic dynamical polarizability, related to the instantaneous dipole moments by $\bmu_{\ell}(k)=\alpha(k) E_{\ell}({\bk},j;\br)$, ${\bf E}({\bk},j;\br )$ the Fourier component of the electric field operator, and $V_{\ell m}(\bk,{\bf R})$ is the usual tensor potential \cite{CT98}.

We now sketch the main elements of our calculation. In order to generalize the method described above to uniformly accelerating atoms \cite{NP13},
we first calculate the electric field due to the accelerating dipole $A$ at the position of the other dipole B, at the retarded time $t_r=t-\rho(t_r)/c$ \cite{PT01}. Here $\rho(t_r)$ is an {\em effective} distance describing the distance traveled by the virtual photons exchanged between the two moving atoms.
For atoms at rest, it coincides with the interatomic distance, that is $\rho(t_r)=R$; if the two atoms are accelerating, their effective distance depends on time because of the time-dependence of the Lorentz factor $\gamma = (1-v^2/c^2)^{-1/2}$. We then calculate and Lorentz-transform the electric and magnetic field generated by the atomic dipoles to the comoving reference frame, where the two atoms are instantaneously at rest.
The interaction energy of the two atoms is thus evaluated in the comoving frame of the accelerating atoms, and finally expressed in terms of quantities measured in the laboratory frame. After some lengthy calculations, assuming a non relativistic motion, we finally get \cite{NP13}:
\begin{eqnarray}
&\ &\Delta \tilde{E}=\Delta E^{\rm r}+\frac{a^2 t}{2 c^3}\frac{\hbar c}{\pi R^3} \int _0^\infty \alpha_A(\rmi u)\alpha_B(\rmi u)
 \left( 3 + \frac{4}{uR} +\frac{2}{u^2R^2}\right)u^2\,\exp[-2uR]\, \rmd u\, \nonumber \\
&\ &+\frac{a^2t^2}{6c^2}\frac{\hbar c}{\pi R^2}\,\int _0^\infty \alpha_A(\rmi u)\alpha_B(\rmi u) \, \left( -1+\frac{4}{uR}+\frac{8}{u^2R^2}+\frac{8}{u^3R^3}+\frac{4}{u^4R^4}\right) u^4\,\exp[-2uR]\,\rmd u
\label{eq:2}
\end{eqnarray}
where $\Delta E^{\rm r}$
is the well-known van der Waals potential energy for atoms at rest, scaling as $R^{-6}$ in the near zone $R\ll\lambda$ ($\lambda$ being the main  atomic transition wavelength) and as $R^{-7}$ in the far zone $R\gg\lambda$  \cite{PT93,CP48}. The other two terms in (\ref{eq:2}) are the changes in the potential energy due to the acceleration.
These two terms depend explicitly on time and are responsible of a qualitative change of the distance-dependence of the interaction energy. In fact, in the near zone a term proportional to $R^{-5}$ adds to the usual $R^{-6}$ term for atoms at rest, while in the far zone a term proportional to $R^{-6}$ adds to the usual $R^{-7}$ behaviour. The presence of the term as $R^{-6}$ in the far-zone is reminiscent of the Casimir-Polder interaction between two atoms at rest at finite temperature $T$ \cite{Barton01}, thus suggesting a connection with the Unruh-thermal effect. Also, the acceleration-dependent corrections are proportional to $a^2t^2$ or $a^2t$. This is a quite relevant feature of our result (\ref{eq:2}), since it suggests that modifications to the interaction energy between the two atoms could be observed also for reasonable values of the acceleration, provided that sufficiently long times are considered \cite{NP13}.
All this could be of relevance for a possible indirect detection of the Unruh effect through the van der Waals force between two accelerated atoms.

\section{Resonance interaction between two accelerated atoms in vacuum}
\label{sec3}

The physical arguments illustrated in the previous section, suggest to investigate if similar effects of acceleration manifest also in other quantum-electrodynamical processes, for example in the resonance interaction between two uniformly accelerated entangled identical atoms. In particular, in this Section we investigate if the effect of a uniform acceleration is equivalent to an Unruh {\em thermal} effect. In this context, it has been recently shown that the Casimir-Polder interaction energy between two uniformly accelerating atoms exhibits a qualitative change from a short-distance thermal behaviour to a long-distance nonthermal behaviour, with respect to a reference length identified with $z_a = c^2/a$, where $a$ is the proper acceleration of the two atoms \cite{MNP14}. In this case the effect is due to the contribution of vacuum fluctuations to the interaction energy. Thus, effects of acceleration beyond the acceleration-temperature equivalence could be expected also in other radiative processes.

We consider here the resonance interaction between two uniformly accelerated identical atoms, one excited and the other in the ground state, prepared in a correlated (symmetric or antisymmetric) state and interacting with the electromagnetic field in the vacuum state. Since resonance interactions are usually much more intense than dispersion  interactions, being a second-order effect, they could be another suitable candidate for observing the effect of accelerated motion in quantum field theory.
Our aim is also to explore physical situations where the effects of the acceleration could be exclusively nonthermal \cite{RLMNSZP16}.
We model the two atoms $A$ and $B$ as two-level systems with energy $\mp\frac{1}{2}\hbar\omega_0$, associated  with the states $\vert g\rangle$ and $\vert e\rangle$, respectively. We assume that the two atoms interact with the electromagnetic field in the vacuum state, and accelerate with the same uniform acceleration along two parallel trajectories in the $x$ direction, $\xAt$ and $\xBt$, maintaining constant their distance. In the comoving frame, their trajectory is  \cite{BD83}
\begin{eqnarray}
&\ &t(\tau)=\frac{c}{a}\sinh\frac{a\tau}{c},\,\,
x_{A/B}(\tau)=\frac{c^2}{a}\cosh\frac{a\tau}{c},\,\,
 y_{A/B}(\tau)=0, \,\,\,z_A(\tau)=z_A,\,\,\,z_B(\tau)=z_B,
\end{eqnarray}
where $\tau$ is the proper time.
We use the multipolar coupling scheme in the dipole approximation; in the instantaneous inertial frame of the two atoms, the Hamiltonian is \cite{CT98, RLMNSZP16}
\begin{eqnarray}
&\ & H= H_{A}+H_{B}+\sum_{{\bf k}j}\hbar \omega_k a^{\dagger}_{{\bf k}j}a_{{\bf k}j}\frac{dt}{d\tau}
- \boldsymbol\mu_{A}(\tau)\cdot{\bf E}(x_{A}(\tau))-\boldsymbol\mu_{B}(\tau)\cdot{\bf E}(x_{B}(\tau)) ,
\label{eq:5}
\end{eqnarray}
where $j=1,2$ is the polarization index, $\boldsymbol\mu=e{\bf r}$ the atomic dipole moment operator, $a_{{\bf k}j}$, $a^{\dagger}_{{\bf k}j}$ the bosonic field operators, and ${\bf E}(x(\tau))$ the electric field operator. $H_{A}$ and $H_{B}$ are the Hamiltonians of the two atoms.

We assume the two identical atoms prepared in their symmetric or antisymmetric state
\begin{equation}
\mid \psi_{\pm} \rangle = \frac 1{\sqrt{2}} \left( \mid g_A, e_B \rangle \pm \mid e_A, g_B\rangle \right) \, .
\label{eq:6}
\end{equation}
In these states the atomic excitation is delocalized between the two atoms. In the Dicke model, the symmetrical (antisymmetric) superposition is the well known superradiant (subradiant) state, because its spontaneous decay is enhanced (inhibited) \cite{Dicke}.

To evaluate the resonance interaction between the two atoms, we exploit a procedure proposed in \cite{DDC82}, which allows to separate at second order in perturbation theory, the contributions of vacuum fluctuations and radiation reaction to the energy shift of the two-atom system.
This approach has been used to investigate the effect of the atomic acceleration on radiative properties of single atoms \cite{Passante98,RS09,AM95,ZYL06,Rizzuto07} and has been also  recently generalized at fourth order in the coupling to investigate dispersion interactions between uniformly accelerated atoms \cite{MNP14,NMZP17}.
Following this procedure, we obtain the resonance interaction between two atoms ($\ell,m = x,y,z$ and the sum over repeated indices is assumed)
\begin{eqnarray}
&\ &\delta E^{res}= -\frac{e^2}{2}\int_{\tau_0}^{\tau}d\tau'\chi_{\ell m}^F(\xAt,\xBtp)C^{A,B}_{\ell m}(\tau,\tau')
+(A \leftrightarrow B \,\mbox{terms}) \, ,
\label{eq:8}
\end{eqnarray}
where $\chi^F(\xt,\xtp)$ and $C^{A,B}(\tau,\tau')$ are the field susceptibility and the atomic symmetric correlation function, respectively \cite{Passante98,DDC82}.
From ({\ref{eq:8}) it follows that the resonance energy shift of the two-atom system, is entirely due to the self-reaction contribution, involving only the field susceptibility, $\chi^F(\xAt,\xBtp)$, and the atomic symmetric correlation function, $C^{A,B}(\tau,\tau')$. The vacuum fluctuations term does not contribute. This is indeed expected on a physical basis, because the resonance interaction originates from an exchange of a real or virtual photons between the two correlated atoms, and therefore it is entirely due to the field radiated by the atoms (source fields). This feature has relevant consequences when we consider accelerating atoms, yielding that, as we now show, resonance interactions do not display any signature of the Unruh {\em thermal} effect.  This situation should be compared with the dispersion interaction between two uncorrelated atoms discussed in the previous section, where the interaction is basically related to the spatial correlations of vacuum fluctuations, and thus thermal effects of acceleration do manifest in the interaction.

Using the appropriate field and atomic statistical functions (susceptibility and symmetric correlation function, respectively), taking the limits $\tau_0\rightarrow -\infty$, $\tau\rightarrow\infty$ in Eq. (\ref{eq:8}), after some algebra we obtain the following expression for the resonance interaction between two accelerated atoms \cite{RLMNSZP16}
\begin{eqnarray}
\delta E^{res}=\pm(\mu^A_{eg})_{\ell}(\mu^B_{ge})_m\biggl\{\frac{1}{z^3}\widetilde{V}_{\ell m}(a, z,\omega_0) +\frac{a}{2z^2c^2} \widetilde{W}_{\ell m}(a, z,\omega_0)\biggr\}\, ,
\label{eq:10}
\end{eqnarray}
where $+$ and $-$ signs refer to the symmetric or antisymmetric state, respectively.

Expression (\ref{eq:10}) is valid for
any value of $az/c^2$, $z = \vert z_B - z_A \vert$ being the interatomic distance.
The functions $\widetilde{V}_{\ell m}(a, z,\omega_0)$ and $\widetilde{W}_{\ell m}(a, z,\omega_0)$, whose explicit expression we do not report here, are a generalization of the stationary tensor potential to the case of accelerated atoms, and  are responsible of the qualitative change of the distance dependence of the resonance interaction energy we are going to discuss (see Ref. \cite{RLMNSZP16} for more details). For $za/c^2\ll 1$, we recover the well-known inertial resonance interaction \cite{CT98}, which
scales as $z^{-3}$  in the near-zone $z\ll\lambda$ and as $z^{-1}$ in the far-zone $z\gg \lambda$ and 
($\lambda =c/\w0$ being the reduced atomic transition wavelength).
On the other hand, at higher orders in $az/c^2$, Eq. (\ref{eq:10}) shows that the atomic acceleration leads to a change of the distance dependence of the resonance interaction, yielding a scaling with a different power law.
For example, if both dipole moments are oriented along the same direction ($x$,$y$, or $z$), it can be shown that $\widetilde{W}_{\ell m}(a, z,\omega_0)=0$, and only the term containing $\widetilde{V}_{\ell m}(a, z,\omega_0)$ contributes in (\ref{eq:10}), and for $za/c^2 \gg 1$ we obtain
\begin{eqnarray}
&\ &\delta E \simeq \pm (\mu_{eg}^{A})_{\ell}(\mu_{ge}^{B})_m \frac{1}{z^3}\Biggl\{(\delta_{\ell m}-q_{\ell}q_m-2n_{\ell}n_m)\biggl[\frac{2\omega_0z}{c} \sin\Bigl(\frac{2\omega_0 c}{a}\ln\biggl(\frac{az}{c^2}\biggr)\Bigl)-\frac{\omega_0^2 z^2}{c^2}\left(\frac{2c^2}{z a}\right)\biggr.\Biggr.
\nonumber\\
&\ &\Biggl.\biggl.\times\cos\Bigl(\frac{2\omega_0 c}{a}\ln\left(\frac{az}{c^2}\right)\Bigr)\biggr]+q_{\ell}q_m\left(\frac{8c^2}{a z}\right)\cos\Bigl(\frac{2\omega_0 c}{a}\ln\biggl(\frac{az}{c^2}\biggr)\Bigr)\Biggr\}\, .
\label{eq:12}
\end{eqnarray}
Here, ${\bf n}=(0,0,1)$ is the unit vector along the $z$ direction and ${\bf q}=(1,0,0)$ is the unit vector along the direction of acceleration, $x$.
On the other hand, when the dipoles are orthogonal to each other, with one of them along $z$ and the other  in the $(x,y)$ plane, only the term with $\widetilde{W}_{\ell m}(a, z,\omega_0)$ contributes to the resonance interaction, being $\widetilde{V}_{\ell m}(a, z,\omega_0)=0$.  Since such a non-diagonal term is present only for $a \neq 0$ and the resonance interaction for stationary atoms vanishes in this specific configuration, its contribution is a sharp signature of the accelerated motion.
Our results clearly show that the non-inertial motion yields a new scaling of the interaction with the distance between the atoms and, since it is not related to vacuum field fluctuations, this qualitative change is not a {\em thermal} effect and goes beyond the usual temperature-acceleration equivalence \cite{RLMNSZP16}.

The nonthermal nature of the effect of acceleration on the resonance interaction induces us to question at which extent the equivalence between acceleration and temperature is valid \cite{ZPR16}. Recent works have concerned with the problem of the equivalence of physical predictions in coaccelerated and comoving frames,
and it was argued that a complete equivalence of the two different frames requires assuming in the Rindler frame an Unruh temperature proportional to the acceleration for the quantum fields \cite{ZY07,MV02}.  These investigations seem to suggest that the existence of the Unruh effect is necessary for the coherence of the results obtained in the comoving and coaccelerated frames.

We now address the question if this assumption is indeed required for all radiative process, or if there is some physical effect for which comoving and coaccelerated observers give the same prediction, without the additional assumption of an Unruh temperature.
We consider the resonance interaction between two accelerated atoms from the point of view of a coaccelerated (Rindler) observer, and discuss the relation with the result \label{eq:9} obtained in the comoving frame.
As before, we suppose the two atoms accelerating with the same acceleration $a$ along the $x$ direction.
The trajectories of the two atoms in the Rindler coordinates $(\tau,\xi,y,z)$ are \cite{BD83}
\begin{eqnarray}
\tau_A=\tau\;,\quad\;\xi_A=y_A=0\;,\quad\;z_A=z_1, \;\quad
\tau_B=\tau\;,\quad\;\xi_B=y_B=0\;,\quad\;z_B=z_2\;.
\label{eq:13}
\end{eqnarray}
Again, we suppose the two atoms interacting with the quantum electromagnetic field in the vacuum state and prepared in one of the correlated states (\ref{eq:6}). Following the scheme in \cite{LOY08}, we quantize the electromagnetic field in the Weyl gauge. In the multipolar coupling scheme and in the dipole approximation, we then calculate the resonance interaction energy following the same procedure developed in the previous section.
From  Eq.(\ref{eq:8}), after lengthy algebraic calculations, we recover the same result given in (\ref{eq:10}) \cite{ZPR16}.
This shows the complete equivalence between the two different, locally inertial and coaccelerated, frames, as far as the resonance interaction is concerned. This agrees with the mentioned fact that the resonance interaction is entirely due to the radiation reaction field, so that nonthermal effects are essentials, while vacuum fluctuations are not relevant. Therefore, although the vacuum in the comoving frame is not equivalent to the vacuum in the coaccelerated  frame, the resonance interaction between the two uniformly accelerated atoms is the same in both frames without the additional assumption of an Unruh temperature for the field in the coaccelerated frame. We conclude that the assumption of an Unruh temperature in a coaccelerated frame is not a requirement for all radiative processes.

\section{Resonance interaction between two accelerated atoms in the presence of a perfectly conducting plate}
\label{sec4}
In this section we briefly illustrate preliminary results on the resonance interaction between two atoms moving with a uniform acceleration parallel to a perfectly conducting plate \cite{ZRP17}. The presence of the reflecting plate changes the field modes and it is of great interest to investigate its effect on the resonance interaction between the two atoms. The atom-wall Casimir-Polder interaction between a uniformly accelerated atom and a perfectly reflecting plate has been already investigated in the literature \cite{RS09, MNPRS14}, and it has been shown that the presence of the boundary affects both vacuum fluctuations and radiation reaction contributions. We thus expect that also the resonance interaction should be modified by the presence of the reflecting plate.

Let us consider two atoms near a perfectly reflecting plate, placed at $z=0$, and interacting with the electromagnetic field in the vacuum state. We suppose the atoms, separated by a distance $L$ along the direction perpendicular to the wall, located at $z_A=z$ and $z_B=z+L$. They have a uniform acceleration along $x$, parallel to the plate. Again, we evaluate the resonance interaction between the two accelerating atoms, given by the radiation reaction contribution, in the presence of a conducting plate. After some algebra and using units such that $\hbar=c=1$, we find \cite{ZRP17}
\begin{eqnarray}
&&\delta E=-\frac{\pi}{4}\biggl[\delta_{\ell m}(\mu^A_{ge})_{\ell}(\mu^B_{eg})_mP_{\ell m}(a,\mathcal{R},\omega_0)+((\mu^A_{ge})_x(\mu^B_{eg})_z+(\mu^A_{ge})_z(\mu^B_{eg})_x)P_{xz}(a,\mathcal{R},\omega_0)\biggr] ,
\label{eq:17}
\end{eqnarray}
where $\mathcal{R}=L+2z$ and $P_{\ell m}(a,\mathcal{R},\omega_0)$ is a function, that we do not report here, which modulates the resonance energy interaction with the distance $\mathcal{R}$ and the atomic acceleration $a$.
Comparing with the case of atoms in free-space previously discussed, we find the emergence of a new  behaviour originating from the presence of the boundary and related to the vectorial nature of the electromagnetic field. In fact, from Eq. (\ref{eq:17}) it follows that the effect of atomic acceleration on the resonance interaction can be enhanced or inhibited through an appropriate choice of the orientation of the two dipole moments, as well as of the distance of the two atoms from the plate. This gives additional possibilities to tailor and control the resonance interaction between accelerated atoms, and it could be also relevant for a possible detection of {\em nonthermal} effects of acceleration.

\section*{Conclusions}
We have discussed and reviewed several physical effects related to the uniform acceleration of a system of two atoms interacting with the electromagnetic field in the vacuum state.
We have first discussed the van der Waals/Casimir-Polder dispersion interaction between two accelerated ground-state atoms, and shown that the accelerated motion yields a new scaling of the interaction with the distance and an explicit time dependence of the interaction between the atoms. Exploiting this result, we have argued that our two-atom system could be a suitable setup for an indirect detection of the Unruh effect, even for a reasonable value of the acceleration. We have then investigated the resonance interaction between two identical atoms, one excited and the other in the ground-state and prepared in a {\em Bell-type} state, moving with uniform acceleration, both in free-space and in the presence of a perfectly reflecting plate. We have shown that the atomic acceleration significantly changes the resonance interaction between the atoms, resulting in a different scaling of the interaction with the distance and in a sharp manifestation of nonthermal effects of the acceleration for specific configurations of the two-atom system.

\ack
W.Z. would like to thank financial support from the National Natural Science Foundation of China (NSFC) under Grants No. 11405091 and China Scholarship Council (CSC).
J.M. acknowledges support from the Alexander von Humboldt Foundation.

\end{document}